\documentstyle[12pt]{article}
\begin{document}
\input psfig.tex

\large

\begin{center}

\bf{An ET Origin for Stratospheric \\
    Particles Collected During the \\
    1998 Leonids Meteor Shower} \\

\vspace{0.25truein}

\normalsize

{David A. Noever$^*$, James A. Phillips$^\dagger$, John M. Horack$^*$, \\
Gregory Jerman$^*$, and Ed Myszka$^\ddagger$} \\

\vspace{0.25truein}

$^*$Science Directorate, NASA/Marshall Space Flight Center, Huntsville, AL 35812
\\
david.noever@msfc.nasa.gov \\
john.horack@msfc.nasa.gov \\
gregory.jerman@msfc.nasa.gov \\
256 544 1872 \\
256 544 5800 (fax)

\vspace{0.25truein}

$^\dagger$BishopWebWorks, 162 Alpine Drive, Bishop, CA, 93514 \\
phillips@spacesciences.com \\
760 873 5585 \\
760 872 1382 (fax)

\vspace{0.25truein}

$^\ddagger$CSC Corporation, SD-01, NASA/Marshall, Huntsville, AL,
35812 \\
ed.myszka@msfc.nasa.gov \\
256 544 1032 \\
256 544 5800

\vspace{0.75truein}

Submitted to {\it Icarus} 10 September 1999
\end{center}

\vfill

\normalsize

\noindent Keywords:  Comets, composition;  Meteors; Interplanetary Dust \\
Number of Figures: 3 \\
Number of Pages: 16 (including Abstract, References, Table, and Figures)

\eject

\baselineskip 0.33truein

\begin{center}
Abstract
\end{center}

\vspace{0.1truein}

On 17 November 1998, a helium-filled weather balloon was launched into the
stratosphere, equipped with a xerogel microparticle collector.  The three-hour 
flight was designed to sample the dust environment in the stratosphere
during the Leonid meteor shower, and possibly to capture Leonid meteoroids.  
Environmental Scanning Electron Microscope analyses of the returned collectors 
revealed the capture of a $\sim$30-$\mu$m particle, with a smooth, 
multigranular shape, and partially melted, translucent rims; similar 
to known Antarctic micrometeorites.  Energy-dispersive X-ray Mass 
Spectroscopy shows enriched concentrations of the non-volatile elements, Mg, 
Al, and Fe.  The particle possesses a high magnesium to iron ratio of 2.96, 
similar to that observed in 1998 Leonids meteors (Borovicka, {\it et al.} 1999)
and sharply higher 
than the ratio expected for typical material from the earth's crust.  A 
statistical nearest-neighbor analysis of the abundance ratios Mg/Si, Al/Si, 
and Fe/Si demonstrates that the particle is most similar in composition to 
cosmic spherules captured during airplane flights through the stratosphere.  
The mineralogical class is consistent with a stony (S) type of silicates, 
olivine [(Mg,Fe)$_{2}$SiO$_{4}$] and pyroxene [(Mg,Fe)SiO$_{3}$]--or oxides, 
herecynite [(Fe, Mg) Al$_{2}$O$_{4}$].  Attribution to the debris stream of 
the Leonids' parent body, comet Tempel-Tuttle, would make it the first such 
material from beyond the orbit of Uranus positively identified on Earth.

\vspace{2.0truein}

Proposed Running Head: {\it ET Origin for Particles Captured During the 1998 
Leonids}

\vfill
\eject

Leonid meteoroids contribute a significant fraction of the annual budget of 
cosmic material falling on Earth (Rietmeijer 1999), 
and numerous groups (e.g., Blanchard {\it et al.} 1969, Maag {\it et al.} 1993)
have attempted to capture and retreive them for analysis.  There have also 
been unintended captures of 
Leonid particles by low Earth orbit (LEO) satellites (Humes and Kinard 1997).
In 1969, 32 partially melted meteoroids were collected (Brownlee and Hodge 1969,
Brownlee {\it et al.} 1997) by  
flying U2 and WB57 aircraft at stratospheric altitudes near 20 km.  These 
meteoroids were members of a general low-level flux of particles that 
produce an average zenithal hourly rate (ZHR) of visual meteors near 6/hr on a
daily basis.  This work has led to a better understanding of 
meteoroid compositions and ablation mechanisms.  However, there has since been 
no systematic effort to sample and study extraterrestrial particles in the 
stratosphere in general, or the Leonids in particular.  The upper stratosphere
remains perhaps the best environment to collect meteor dust, however success 
in doing so is a compromise between environmental and technical factors
(Rietmeijer 1999).
 
On 17 November 1998 we launched a 10m weather balloon equipped with an xerogel 
dust collector to sample the stratosphere during the peak flux of the Leonids.  
A matrix of low-density silica xerogels in separated polystyrene wells were 
fixed to the outside of the balloon package.  The payload was carried to an 
altitude above 98$\%$ of Earth's atmosphere during a 1.9 hr flight.  An on 
board digital camera captured eight Leonid fireballs brighter than magntitude 
-10.  At its maximum altitude, the balloon ruptured as planned and the 
payload descended to Earth by parachute.  Eight candidate impactors were 
analyzed in the returned payload.  One of these exhibits chemical and 
morphological signatures indicating extraterrestrial origin.

Twenty-four one-inch-diameter circular wells of xerogel were sent aloft to the 
stratosphere and only a few were damaged during the balloon's 
descent and landing.
A visual microscopic survey of the remaining xerogel containers revealed that 
all were pitted with craters in the 20 - 100 micron range (e.g., Figure 1).  
Based on the apparent density of craters in each capture well, we selected one 
1-inch diameter circular sample for further study with an Environmental 
Scanning Electron Microscope (ESEM) (Danilatos {\it et al.} 1982).  
The ESEM, which does not require hard vacuum 
conditions to reduce scattering, is normally used to study wet, biological 
samples.  Its advantage for the Leonids sample return is that it is unlikely
to cause vacuum damage to fragile microparticles.
The ESEM was equipped with a 20 keV energy dispersive X-ray mass spectrometer 
(EDS) sensitive to elements with atomic weights $Z$ $>$ 10.  Using the SEM and 
EDS together, we can image each impactor while simultaneously characterizing 
its elemental composition.

We scanned the xerogel sample to a surface depth of 1-2 micron
using a rastered beam similar in size to the diameter of the impactors
(10 - 10,000x, sample-dependent 4 nm resolution).  These instrument
settings were selected to avoid sources of error involving resolution and 
diffraction effects.  ESEM-EDS analysis of the xerogel capture media revealed 
a smooth silica surface to the resolution of approximately 1 micron and 
chemical analysis showed Si:O ratios between 2:1 and 3:1, as expected.  
Residual levels of Fe, Mg, Ni, Ca, Al and C were below 1$\%$ detection 
thresholds.

The eight crater-like pits were examined in the same fashion as the xerogel 
background. Each crater contained a single impactor ranging in size from 
1 - 40 microns.  These particles fell into two categories differentiated by 
morphology and chemical composition.

Seven particles were similar; spherical in shape (Figure 2a) with a strong
signature of Si.  The dominance of Si in the spectra of these 
particles makes accurate abundance measurements difficult because of possible 
confusion with the Si-rich xerogel capture media.  The shape of the particles 
suggests that they have experienced sufficient heating -- possibly a result of 
atmospheric friction -- to render them molten before cooling and reforming as a
sphere.  Without more reliable abundance measurements it is impossible to say
whether or not these seven impactors have an extraterrestrial origin.

An eighth candidate stood out as distinct from the others.  EDS data 
(Figure 2b) showed that this impactor is rich in Si, but also 
has significant concentrations of Mg, Al, and Fe.  The particle is 
irregular in shape with translucent rims and an opaque core, much like known 
cosmic spherules (Brownlee {\it et al.} 1997).  
There is no sign of bulk melting.  The general 
morphology of this 30 $\mu$m particle is also similar to that of Antarctic 
micrometeorites composed of silicates in the 50 - 100 $\mu$m range (Genge {\it
et al.} 1996).

No degassing vesicles or gas corrosion 
from volitiles are apparent as might be expected for intensely heated particles.
It is nevertheless rich in the non-volatile elements Mg, Al, Si and Fe 
(see Fig 2C).   Abundance ratios of Mg, Al, Si and to a lesser extent Fe have 
previously served (Brownlee and Hodge 1969, Love and Brownlee 1991, Love and
Keil 1995) for identification of cosmic particles in three 
broad categories: a dominant type S (stony), and less frequent types I
(iron) and FSN (iron-nickel-sulfur).  S-types, which are thought
to arise from asteroids, dominate the background terrestrial flux.
In stratospheric collections of microparticles, FSN types feature
prominantly but rarely exceed 20 $\mu$m in diameter.  The strong
Si peak in all the EDS analyses (Figures 2a,b) excludes this particle 
from the I-spherules which contain no silicates.

The ratios Al/Si, Mg/Si and Fe/Si obtained for the irregular
Leonid meteoroid candidate appear in the first three data columns of
Table 1 along with average ratios for known cosmic spherules, meteorites, 
and typical terrestrial dust.  Also tabulated are the ratios for an xerogel 
control sample which remained on Earth during the balloon flight.  The 
composition of the Leonid candidate is clearly similar to that of cosmic 
spherules previously found on Earth, and much less like that of Earth dust or
the control.  

A special subclass of vitreous S-type cosmic spherules have previously been
detected and identified in stratospheric collections (Brownlee and Hodge 1969,
Brownlee {\it et al.} 1997). 
Members of the subclass smaller
than $\sim$30 $\mu$m were enriched in Al and depleted in Fe relative
to parent chondritic material.  This Leonid candidate may be related
to these particles, as it is also rich in aluminum (Al/Si=0.56),
and slightly depleted in magnesium (Mg/Si=0.89) and iron (Fe/Si=0.30)
compared to median cosmic spherules from deep sea, Antactic, and
stratospheric collections (Brownlee {\it et al.} 1997).

Also notable is the Leonid candidate's high magnesium to iron
ratio.  Airborne UV/Vis spectroscopy of a bright
1998 Leonids meteors show atomic metal spectral lines indicating
Mg/Fe$\approx$3.3 (Borovicka {\it et al.} 1999).  This is in good agreement with the high Mg/Fe ratio
of the Leonid sample (2.96) and in sharp contrast to the low ratio expected
for typical material from the earth's crust (Mg/Fe $\sim$ 0.4).

Iron/Nickel ratios can effectively
discriminate between terrestrial and extraterrestrial 
origin of stratospheric dust (Rietmeijer 1999).  Terrestrial dust particles 
tend to have high ratios.  For example, volcanic ash from Mt. 
St. Helens collected at 34-36 km altitude had Fe/Ni=1200.  
Extraterrestrial dust exhibits much lower ratios, e.g., optical 
spectra of meteors yield Fe/Ni $\sim$ 19.  Nickel was not detected in the 
EDS spectrum of the Leonids candidate, as the Ni peak was below the 
3$\sigma$ noise level of 1$\%$ fractional composition.  The Fe peak was 
measured at 9$\%$, placing a lower limit of Fe/Ni $>$ 9.  This test does not 
yield discrimination between terrestrial and extraterrestrial dust.  

To test the extraterrestrial hypothesis further, we use a 
statistical procedure to compare EDS data with the chemical makeup of cosmic 
microparticles in various sampling databases (Rietmeijer 1999, Brownlee {\it et
al.} 1997).   We express compositional ratios as a vector $X$ with 
components (Mg/Si, Al/Si, Fe/Si). The `chemical distance' $R_{Lj}$ between the 
Leonid candidate ($L$) and any other material ($X_j$) may then be evaluated as 
$$ R_{Lj} = {\sqrt{\sum_{i= 1}^3  (L_i-X_{ij})^2}}. \eqno(1)$$  
This formulation allows us to compare and classify objectively the Leonid 
candidate with nearest neighbors based on their minimum Euclidian distance.  
These data are contained in the fourth column of Table 1, and the ``chemical
distance'' vectors are plotted in Figure 3.

These data clearly indicate that the 1998 Leonid candidate is
closest in composition to coarse unmelted cosmic spherules
captured during previous airplane flights through the
stratosphere.  The low concentration of iron in those
spherules, relative to the Leonid candidate, may result from
volatilization of iron sulfide during atmospheric
entry or perhaps unique chemical features for Leonids-related dust
streams.  While the composition of the Leonid candidate
does not perfectly match that of any catalogued cosmic spherule, the 
chemical distance is relatively small.  Importantly, the Leonid particle is 
significantly more distant from the control than the cosmic particles, and 
lies {\it most} distant from terrestrial dust.

This analysis shows conclusively that the Leonid candidate is most chemically 
similar to known extraterrestrial particles.  Both its composition and 
morphology are consistent with that of micrometeoroids previously 
gathered from the stratosphere and elsewhere.  Taken together, we believe
the chemical and morphological properties are most consistent with the 
explanation of an extraterrestrial origin. 

However, there are factors which may contravene these conclusions.
It is difficult to estimate {\it a priori} capture probabilities relevant to 
this experiment, as little is known about the stratospheric particle 
environment during an intense meteor shower.  For satellite impacts during
the 1998 shower, a flux of 0.4-59 m$^{-2}$ 
hr$^{-1}$ was predicted,$^7$ corresponding to an expected maximum of 
$\sim$2 impacts/hr for the $\sim$600 cm$^2$ hr area-time product for this 
flight.  While comparable to that observed, the figure is unlikely to be 
meaningful in this context.  The `true' expected value at 20 km depends on a 
very uncertain extrapolation of meteoroid momenta and densities from the top 
of Earth's atmosphere downward to the stratosphere.

The xerogel dust collector was exposed for the entire duration of the balloon 
flight from launch to landing.  We cannot exclude the possibility that 
metal-rich contaminants such as volcanic dust or industrial pollutants were 
captured at low altitudes.  Small ($<$ 10-15 $\mu$m) volcanic ash particles
have been captured in the upper stratosphere (Rietmeijer 1993, Testa {\it et
al.} 1990), and some terrestrial 
particles, albeit atypical ones, are found to have high Al content as does 
the Leonids candidate.  Future 
balloon flights, including one scheduled for the 1999 
Leonids meteor shower, will carry a remotely controllable sample collector 
that opens only while the balloon is in the stratosphere.

Meteoroids sampled in the stratosphere enter the atmosphere at high
speed, $\sim$70 km/s for the Leonids.  Terminal velocity for cometary debris 
in the 20 - 70 micron range is widely thought to be reached at altitudes 
considerably higher than 20 km, thus requiring a significant `drift time' to
reach these lower altitudes.  Empirical data has little to say on this 
point, however, simply because there has been no systematic sampling of 
stratospheric meteoroid fluxes during major meteor showers prior to 1998.  If 
the Tempel-Tuttle debris stream includes a component of larger-, harder-, and 
faster-than-average meteoroids, then this balloon would be sensitive to 
Leonids in real-time during the shower's peak, rather than older, 
slower-moving particles.

Even if the particle is extraterrestrial, it is still possible that it did not 
arise from the debris stream of comet Tempel-Tuttle.  
Most meteoroids do not plunge directly to Earth after entering the atmosphere.
Instead they lose much of their kinetic energy high above the stratosphere and 
very slowly drift downward.  If this scenerio applies to the particles 
captured during the peak of the Leonids meteor shower, they may
have entered Earth's atmosphere days or weeks earlier.   The
irregular particle might have originated, for example, from 
the debris stream of comet Giacobini-Zinner which caused an intense meteor 
shower in October 1998.  Alternatively, it could be a member of the 
low-level background population of meteoroids that permeates the inner solar 
system.  Indeed, the SEM survey of the xerogel collector was a surface scan
and, thus, preferentially sensitive to older, lower velocity impactors.

This year our group will execute a systematic campaign of
balloon flights during relatively intense meteor showers and
also during periods of low meteor activity to evaluate further the
temporal correlation between visual meteor counts and
meteoroid flux in the stratosphere.   These additional flights may
provide the information needed to confirm the present candidate as a Leonid 
or to assign it to a different meteoroid population.

\vfill
\eject

\begin{center}
{\bf REFERENCES}
\end{center}

\begin{enumerate}

\item Blanchard, M. B., Ferry, G. V., $\&$ Farlow, N. H., 1969, {\it Meteoritics},
{\bf 4,} 152.

\item Borovicka, J., Stork, R., $\&$ Bocek, J., 1999 {\it Meteoritics}, (in
press).

\item Brownlee, D. E., $\&$ Hodge, P. W., 1969, {\it Meteoritics}, {\bf 4,} 264.

\item Brownlee, D. E., Bates, B., $\&$ Schramm L., 1997, {\it Meteorit. and Plan.
Sci.}, {\bf 32,} 157-175.

\item Danilatos, G. D., $\&$ Postale, R., 1982, {\it Scanning Electron Microscopy,}
1-16.

\item Genge, M. J., Hutchison, R., $\&$ Grady, M. M., 1996, {\it Meteorit. and Plan.
Sci.}, {\bf 31,} (suppl), A49.

\item Humes, D. H., and Kinard, W. H., 1997, Hubble Space Telescope Archive,
http://setas-www.larc.nasa.gov/HUBBLE/PRESENTATIONS/ \\
hubble$\_$talk$\_$humes$\_$kinard.html

\item Love, S. G., $\&$ Brownlee, D. E., 1991, {\it Icarus}, {\bf 89}, 26-43.

\item Love, S. G., $\&$ Keil, K., 1995, {\it Meteoritics}, {\bf 30,} 269-278.

\item Maag, C. R., Tanner, W. G., Stevenson, T. J., Borg, J., Bibring, J.-P., 
Alexander, W. M. $\&$ Maag, A. J., 1993, in {\it Proc. 1$^{st}$ European Conference on Space 
Debris}, Darmstadt, Germany,
125-130.

\item Rietmeijer, F.J.M.,{\it J. Volc. Geothermal Res.}, 1993, {\bf 55,}
69-83.

\item Rietmeijer, F.J.M., 1999, {\it 37$^{th}$ AIAA Aerospace Sciences Meeting and
Exhibit}, AIAA {\bf 99,} 0502.

\item Testa, J. P., Stephens, J. R., Berg, W. W., Cahill, T. A., Onaka, T.,
Nakada, Y., Arnold, J. R., Fong, N., $\&$ Sperry, P. D., 1990, {\it Earth
Planet. Sci. Lett.}, {\bf 98,} 287-302.
%
%
\end{enumerate}

\vfill
\eject

\begin{center}

{\bf TABLE I.}

\vspace{0.5truein}

\begin{tabular}{| l | r | r | r | c | r |}
\hline
	      &      Mg/Si  &    Al/Si  &  Fe/Si  & & $R_{Lj}$ \\
\hline
	     
{\bf 1998 Leonids}  &		    &      	&       & &	\\
   Sample     &     0.89    &   0.56    &	0.3 	& & 0.000	\\
   Control    &     0.0     &   0.0     &       0.0	& & 1.094 \\

\hline

{\bf Sampling Site}	&	    &		&	& &	\\
  Stratosphere  &  1.06     &   0.233   &  0.633	& & 0.497 \\
  Antarctic     &  1.06     &   0.091   &  0.528        & & 0.549 \\
  Deep Sea      &  1.06     &   0.083   &  1.024        & & 0.884 \\
  All		&  1.06	    &	0.094	&  0.937	& & 0.807 \\

\hline

{\bf Stratospheric unmelted}	&	&	&	& &	\\
  Smooth        &  0.82     &   0.082   & 0.742		& & 0.655 \\
  Porous        &  1.02     &   0.07    & 0.705		& & 0.649 \\
  Coarse        &  1.2      &   0.075   & 0.585         & & 0.642 \\
 Bulk IDP       &  0.98     &   0.075   & 1.08		& & 0.923 \\

\hline

{\bf Bulk Chondrites}	&	    &		& 	& &	\\
    CI          &  1.07     &  0.085    & 0.9		& & 0.786 \\
    CM          &  1.05     &  0.095    & 0.819		& & 0.715 \\
    H           &  0.96     &  0.07     & 0.818		& & 0.717 \\
    L           &  0.93     &  0.069    & 0.584		& & 0.569 \\

\hline

{\bf Earth dust}$^*$   &    0.83   &    $<$ 0.003 &   2.27  &  & 2.048	\\
\hline
\end{tabular}

\vspace{0.25truein}

$^*$values for Earth dust are based on these mass fractional
abundances for Earth as a whole:  34.6$\%$ Fe, 29.5$\%$ O, 15.2$\%$ Si,
12.7$\%$ Mg, 2.4$\%$ Ni,  1.9$\%$ S, 0.05$\%$ Ti. 

\end{center}

\vspace{0.5truein}

{\it TABLE I. Comparison of chemical elemental ratios for non-volatiles in
the Leonids candidate particle, along with known cosmic dust, control sample,
and terrestrial composition.  $R_{Lj}$ expresses the ``chemical distance''
between the Leonids candidate and other particles as described in the text.
The Leonids candidate is most distant from terrestrial composition, and agrees
most closely with known cosmic dust.}

\eject


\begin{center}
{\bf FIGURE CAPTIONS}
\end{center}

\vspace{1.0truein}

{\it FIGURE 1: Sequence showing the cross-section of a particle and impact
crater in the xerogel collector.  The crater measures 20-30$\mu$m in diameter.}
\vspace{0.5truein}

{\it FIGURE 2a: Representative EDS mass spectrogram and ESEM image
of a spherical particle found embedded in the xerogel collector.}
\vspace{0.5truein}

{\it FIGURE 2b: EDS mass spectrum of the irregular, non-volitile rich
particle with translucent rims and an opaque core, and its ESEM image.  
The percentage composition of the sample is Si (31$\%$), Mg (28$\%$), 
Al (18$\%$), O (12$\%$) and Fe (9$\%$) with no appreciable Ni or C.}
\vspace{0.5truein}

{\it FIGURE 3: Three dimensional scatter plot of the ``Chemical Vectors''  
(Fe/Si, Al/Si, Mg/Si) for known cosmic particles (square), the leonids ET
candidate (filled circle) and for terrestrial dust (triangle).  Projections in
the XY plane are also shown to aid in visualization.  The Leonids sample 
particle lies most closely to known particles of extraterrestrial origin, 
and most distant from terrestrial composition.}

\vfill
\eject

\begin{center}
{\bf FIGURES}
\end{center}

\begin{center}
\psfig{figure=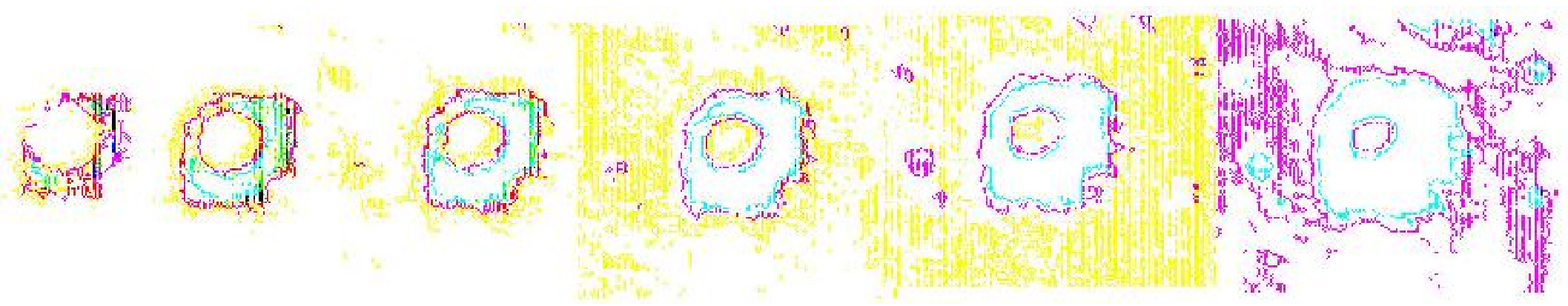,width=6.0in}
\end{center}

\vfill

Figure 1, Noever, et al.; {\it An ET Origin \dots}

\eject

\begin{center}
\psfig{figure=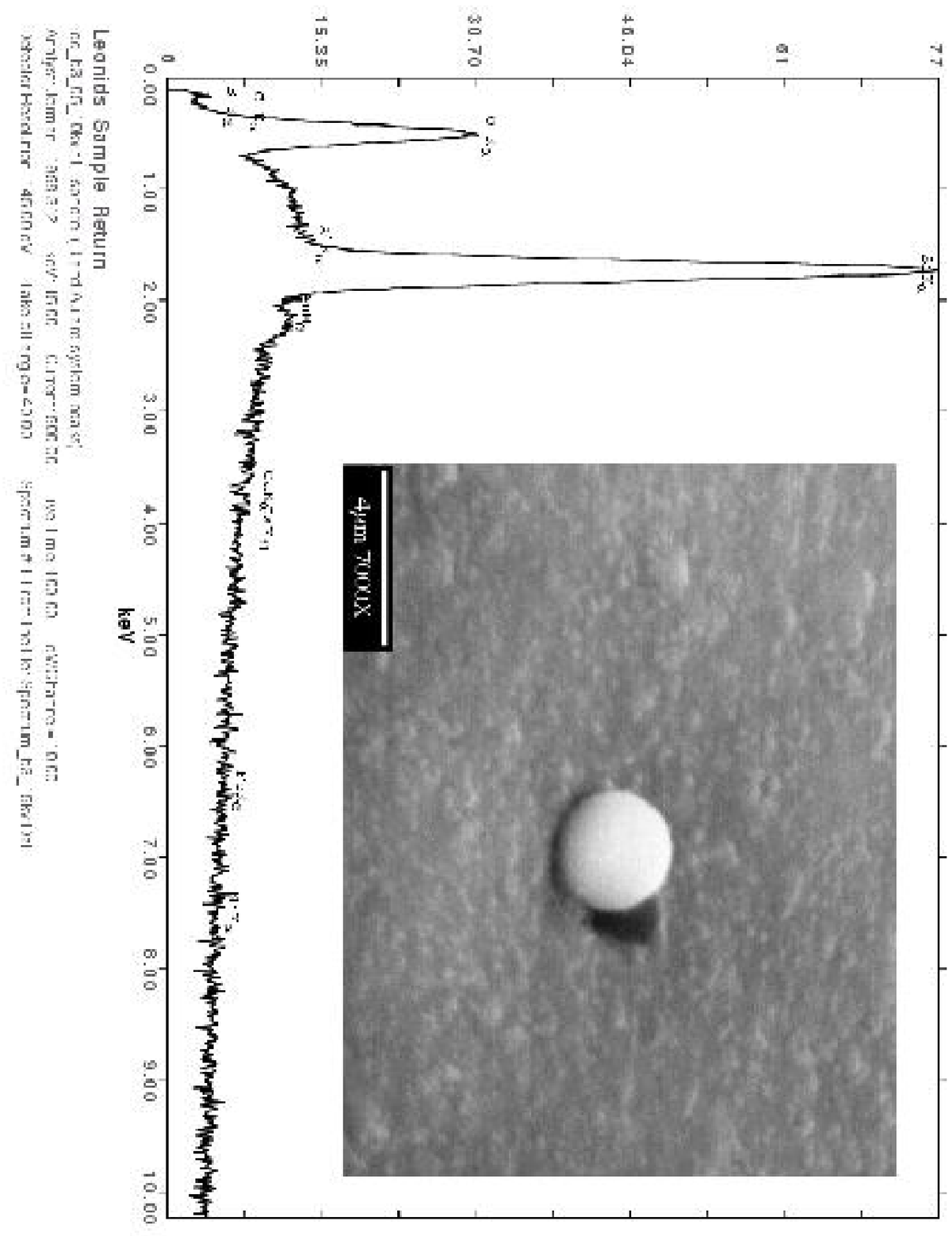,height=7.5in}
\end{center}

\vfill

Figure 2a, Noever, et al.; {\it An ET Origin \dots}

\eject

\begin{center}
\psfig{figure=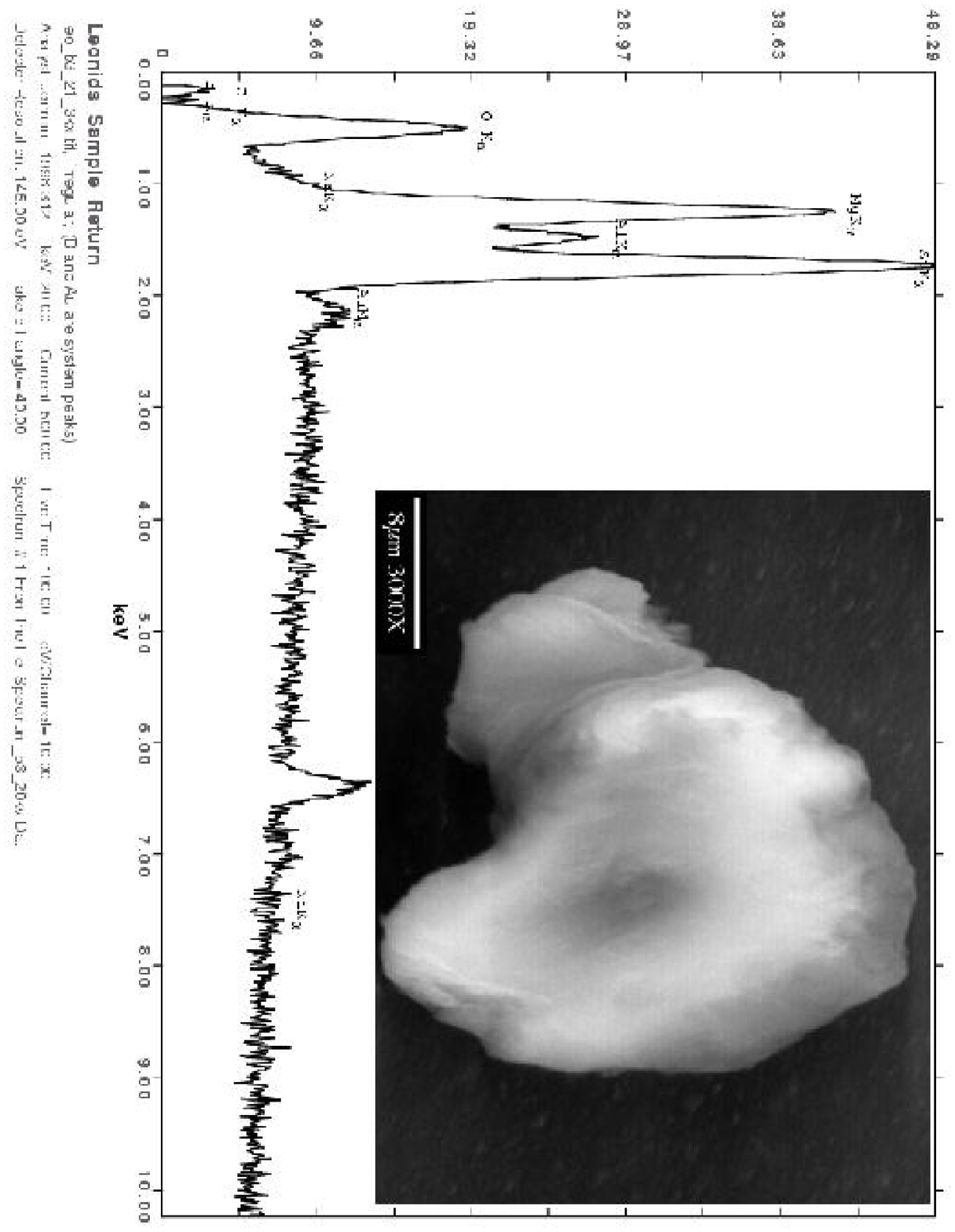,height=7.5in}
\end{center}

\vfill

Figure 2b, Noever, et al.; {\it An ET Origin \dots}

\eject

\psfig{figure=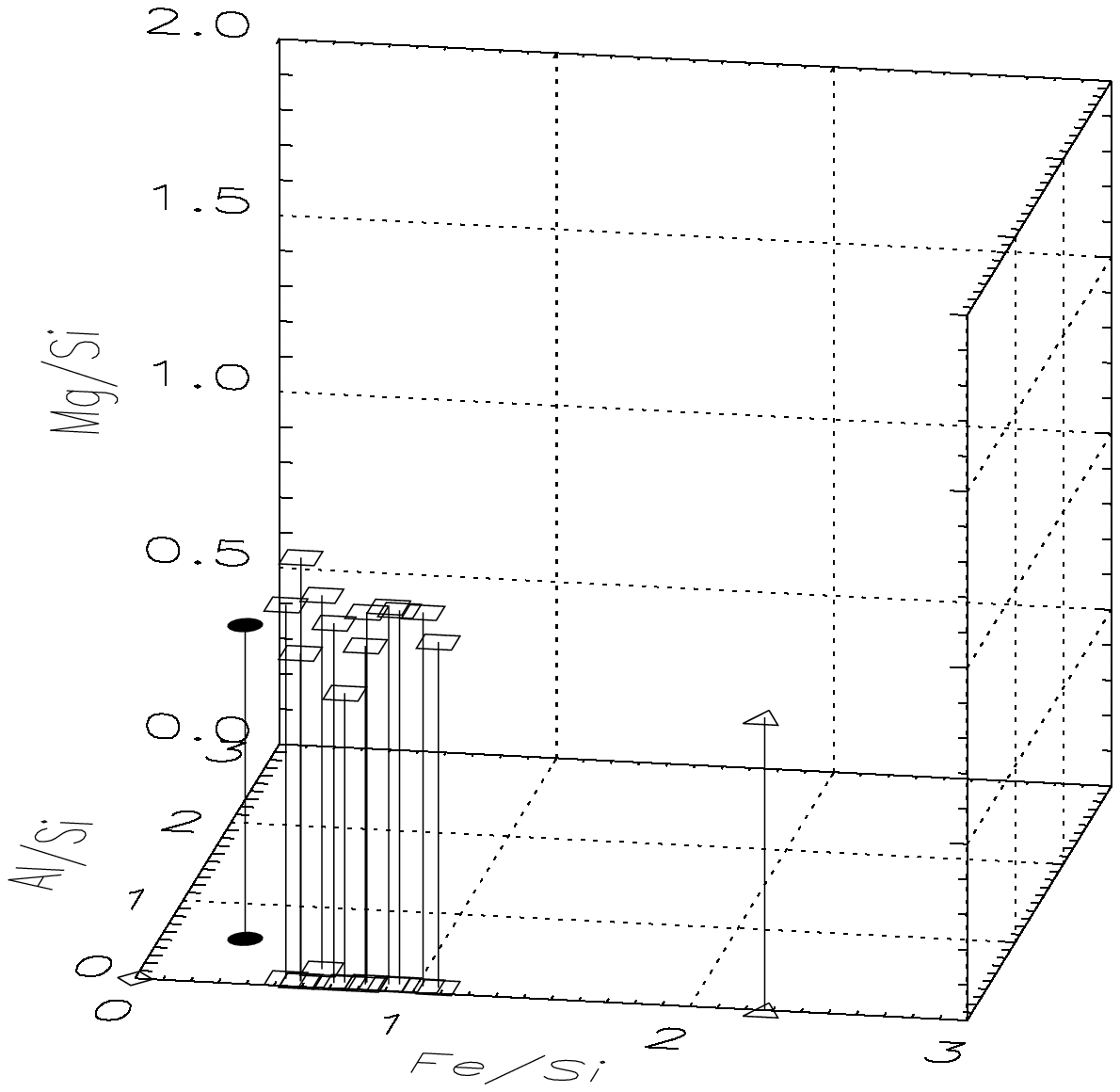}

\vfill

Figure 3, Noever, et al.; {\it An ET Origin \dots}

\end{document}